\documentclass[sigconf]{acmart}
\usepackage{placeins}

\usepackage{colortbl}
\usepackage{nth}
\usepackage{capt-of}
\usepackage{comment}
\usepackage{multirow}
\usepackage{float}
\usepackage{booktabs}
\usepackage{enumitem}
\usepackage{multicol}
\usepackage{graphicx}
\usepackage{subcaption}
\usepackage{amsmath}

\usepackage{arydshln}

\newcommand{\NUMARTICLES}{184736~}
\newcommand{\OUREMBEDDING}{NT2V~}

\setcopyright{acmcopyright}
\copyrightyear{2021}
\acmYear{2021}

\acmConference[Preprint '21]{Preprint}{}{}

\begin{document}

\title{Tell Me Who Your Friends Are: Using Content Sharing Behavior for News Source Veracity Detection}


\author{Maur\'{i}cio Gruppi}
\email{gouvem@rpi.edu}
\affiliation{%
  \institution{Department of Computer Science}
  \institution{Rensselaer Polytechnic Institute}
  \city{Troy}
  \state{NY}
  \country{USA}
  \postcode{12180}
}

\author{Benjamin D. Horne}
\email{bhorne6@utk.edu}
\affiliation{%
  \institution{School of Information Sciences}
  \institution{University of Tennessee Knoxville}
  \city{Knoxville}
  \state{TN}
  \country{USA}
  \postcode{37996}
}

\author{Sibel Adal{\i}}
\email{adalis@rpi.edu}
\affiliation{%
 \institution{Department of Computer Science}
  \institution{Rensselaer Polytechnic Institute}
  \city{Troy}
  \state{NY}
  \country{USA}
  \postcode{12180}
 }

\begin{abstract}
Stopping the malicious spread and production of false and misleading news has become a top priority for researchers. Due to this prevalence, many automated methods for detecting low quality information have been introduced. The majority of these methods have used article-level features, such as their writing style, to detect veracity. While writing style models have been shown to work well in lab-settings, there are concerns of generalizability and robustness. In this paper, we begin to address these concerns by proposing a novel and robust news veracity detection model that uses the content sharing behavior of news sources formulated as a network. We represent these content sharing networks (CSN) using a deep walk based method for embedding graphs that accounts for similarity in both the network space and the article text space. We show that state of the art writing style and CSN features make diverse mistakes when predicting, meaning that they both play different roles in the classification task. Moreover, we show that the addition of CSN features increases the accuracy of writing style models, boosting accuracy as much as 14\% when using Random Forests. Similarly, we show that the combination of hand-crafted article-level features and CSN features is robust to concept drift, performing consistently well over a 10-month time frame. 
\end{abstract}



\maketitle

\section{Introduction}
The spread of false and misleading news is damaging to society~\cite{lewandowsky2012misinformation,lazer2018science}. Its harms can be felt across many parts of society, including
politics~\cite{allcott2017social},
education~\cite{alvermann2017social}, and health~\cite{speed2017rise,singh2020first,pennycook2020fighting}. Due to this cost, limiting false and misleading news has become a concern for both researchers and practitioners.

Due to the scale of this problem, many researchers have built
classifiers to automatically assess the veracity of
news~\cite{kumar2018false}. The vast majority of these newly-developed
classifiers are based on features of the text in news articles or
claims~\cite{baly2019multi,potthast2017stylometric,popat2016credibility,horne2020all}. These text-based methods have been shown to
work well in lab-settings because unreliable news is often
written in a different style than reliable news, employing many
different linguistic and grammatical markers. These differences are often attributed to various factors, such as the use of moral-emotional language to gain engagement~\cite{brady2017emotion}. Despite this success, there are still
concerns about the robustness of these methods. Specifically, text-based methods are prone to performance degradation over time (often called \textit{concept drift})
due to the dynamic attributes of the news cycle~\cite{horne2019robust}. Furthermore, text-based models may be dependent on language or over-fit to specific domains or topics, making them less generalizable.

In this paper, we present an alternative and complementary method for detecting unreliable information based on the behavior of news producers. Specifically, past work has documented that many news producers copy news stories from each other. In essence, copying is a type of amplification, making a story available to the readers of a specific source. In mainstream media, this has been attributed to meeting the demand of all-day news consumption~\cite{boczkowski2010news}. However, this behavior is very common in alternative media as well, with different motivations. These motivations include generating engagement at a low cost, increasing perceived credibility of stories, and their algorithmic visibility in social media platforms. It has been shown that when this behavior is formulated as a network, the community structures found in the network correspond to different types of news sources in the media ecosystem, including mainstream media, hyper-partisan media, and more~\cite{starbird2017examining,horne2019different}.


Building on the pervasive nature of content sharing among news producers, we propose a new set of source veracity features using content sharing networks, or `CSN' for short. Our hypothesis is that the network location of sources in the CSN can provide a strong signal of source reliability. We introduce three feature sets using CSNs, one set based on well-known network properties of nodes and two sets using network embedding methods. To show the effectiveness of CSN features, we conduct a thorough study comparing them to previously studied text-based features. Furthermore, we test the stability of different models to over time. Through our comprehensive study, we show that CSN information alone outperforms the previously used text-based methods. Despite the high accuracy of CSN-only models, the combination of CSN information and text information works best, increasing accuracy by at most 14.7\% over text-only models. We also show that the combination of CSN models and text-based models provide stable performance over time. Additionally, we find that text and CSN models are highly complementary: they make different types of errors in our data set. The text models make fewer errors when predicting reliable sources and the CSN models make fewer errors when predicting unreliable sources.

In short, using the content sharing behavior of news sources in 
veracity detection leads to highly accurate models. By adding complementary information to existing text models, we improve the overall performance and enhance model robustness.

\section{Related Work}

There is a large body of work on news veracity detection, particularly focused on political news articles since 2016~\cite{kumar2018false}. These works have used a variety of machine learning techniques. These techniques include binary supervised models~\cite{baly2018,horne2018accessing,horne2019robust,castelo2019topic,cruz2019team}, multi-class supervised models~\cite{baly2019multi}, semi-supervised models~\cite{guacho2018semi,agerri2019doris}, unsupervised models~\cite{hosseinimotlagh2018unsupervised}, and various Neural Network models~\cite{singhania20173han,li2019encoding,farber2019team,moreno2019rouletabille}. Some works have also framed the problem as a ranking problem, rather than a classification problem~\cite{ye2019mediarank,barron2019proppy}. The primary features of these detection methods are based on the article text, many of which are hand-crafted feature sets. These text features range from very specific, such as the bias and emotion in an article, to very generic, such as the term frequency within an article. In general, these types of features have been shown to work well and can be used to explain algorithm decisions, but they are prone to sub-optimal performance over time and across domains. Theoretically, they are also prone to text manipulation from malicious sources~\cite{horne2019robust}, although this behavior has not yet been shown in real life.

One method of strengthening these text-based models is to augment them with features unrelated to the content of the article. To some degree, this has been done. Baly et al. add the presence of a Wikipedia page and Twitter account for each source~\cite{baly2018} to article-related feature models. Similarly, Li and Goldwasser use both text features and Twitter social features to detect veracity~\cite{li2019encoding}. Ye and Skiena add the number of advertisements on a page and the popularity of the source to text-based ranking models~\cite{ye2019mediarank}. Castelo et al. add various web markup features such as the presence of an article author, number of advertisements, and number of images~\cite{castelo2019topic}. However, with the exception of the number of advertisements, these additional features can be easily manipulated with little cost to the malicious news producer. 

Mixing text features with source-level features has also been done in false claim and rumor detection (rather than news article or news source veracity detection). Many studies of false claims on Twitter utilize features of the users who spread the claim, such as number of followers, number of friends, age of profile, or temporal patterns of the user posts~\cite{castillo2011information,ruchansky2017csi,yang2012automatic}. Other claim veracity works have used popularity as a feature~\cite{popat2016credibility}. Again, these additional non-text features are shallow and easy to manipulate. 

In this paper, we address this gap by introducing a new source-level, behavioral feature for the news source veracity prediction task, namely content sharing behavior. This behavior is costly to manipulate and highly consistent over time, which lends itself to building robust prediction models for the task. This cost stems from the additional effort malicious news producers would need to exert to produce independent false content by not copying content from their peers. Further discussion of network construction and the intuition behind using content sharing networks as signals of veracity can be found in Section~\ref{sec:network}. 




\section{Data}\label{sec:data}
In this work, \textbf{given a news article from an unknown source, our goal is to predict if the source of the article is reliable or unreliable}. To this end, we extract news article data from the NELA-GT-2018\footnote{\url{dataverse.harvard.edu/dataverse/nela}} data set~\cite{norregaard2019nela}. The NELA-GT-2018 data set is a political news data set that contains 713K articles from 194 sources, containing all articles by these sources from February 1st, 2018 to November 30th, 2018. These sources come from a wide range of mainstream and alternative media, including many conspiracy-spreading news sources and hyper-partisan blogs. Included in the NELA-GT-2018 data set are source-level labels of credibility from several assessment platforms. Two of the assessment platforms will be used for labeling sources in this paper: Open Sources and NewsGuard\footnote{\url{http://www.newsguardtech.com/}}. Open Sources ratings have been used in many other studies. It uses a panel of experts to mark sources as one or more of these 13 categories: \textit{reliable, blog, clickbait, rumor, fake, unreliable, biased, conspiracy, hate speech, junk science, political, satire,} and \textit{state news}. The criteria for deciding source labels on Open Sources is available on their website. NewsGuard is an independent journalistic organization that similarly uses a group of experts to score news sources based on credibility and transparency using a stringently developed rating process. Specifically, NewsGuard rates sources on the following criteria, with each criteria having an assigned weight:
\begin{enumerate}
    \item Does not repeatedly publish false content (22 points)
    \item Gathers and presents information responsibly (18 points)
    \item Regularly corrects or clarifies errors (12.5 points)
    \item Handles the difference between news and opinion responsibly (12.5 points)
    \item Avoids deceptive headlines (10 points)
    \item Website discloses ownership and financing (7.5 points)
    \item Clearly labels advertising (7.5 points)
    \item Reveals who's in charge, including any possible conflicts of interest (5 points)
    \item Provides information about content creators (5 points)
\end{enumerate}

Using these two sets of source-level labels, we create two classes of news: \texttt{reliable} and \texttt{unreliable} as follows. We extract all articles from sources that have a credibility score above 90 according to NewsGuard to create our reliable class and sources that have a credibility score below 40 or sources that are marked as unreliable/conspiracy/fake by Open Sources to create our unreliable class. Often sources with a score below 40 by NewsGuard are also marked as unreliable/conspiracy/fake in Open Sources. To obtain a score above 90 by NewsGuard, a source would only be allowed to miss one of the last four criteria (criteria 6, 7, 8, or 9). 


Based on this labeling method, we extract \textbf{\NUMARTICLES articles} from \textbf{52 sources}, where 25 sources are marked as reliable and 27 are marked as unreliable. These articles cover 10 months in 2018 (February through November). The sources in each class can be found in Table~\ref{tbl:sources}. 



\begin{table}[ht!]
\centering
\fontsize{9.5pt}{9.5pt}
\selectfont
\begin{tabular}{cc}
\textbf{(R) Reliable sources} & \textbf{(UR) Unreliable sources}\\
\midrule
Reuters & True Pundit\\ 
NPR & Natural News \\
USA Today & Infowars \\
CNN & Veterans Today \\
The New York Times & Activist Post \\
CBS News & Mint Press News \\
WSJ Washington Wire  & Waking Times \\
The Hill & Intellihub \\
CNBC & NODISINFO \\
PBS & TheAntiMedia\\
The Guardian & Freedom Daily \\
Politico & FrontPage Magazine \\
The Denver Post& Conservative Tree House \\
BBC & Shareblue \\
Business Insider & Bipartisan Report \\
Washington Examiner & Newswars \\
Yahoo News & Prison Planet \\
The Daily Beast & The Gateway Pundit \\
Real Clear Politics & Pamela Geller Report \\
National Review & Western Journal \\
New Yorker & The Political Insider \\
Fortune & The Duran \\
Newsweek & Instapundit \\
Mercury News & Palmer Report \\
The Atlantic & Freedom Outpost \\
& The Right Scoop \\\bottomrule
&
\end{tabular}
\caption{Sources used in each class. Note, for BBC we only extract article from their U.S. news feed. These labels are based on external labeling. See Section~\ref{sec:data} for more details.}
\label{tbl:sources}
\end{table}

\section{Using Content Sharing Networks as a Signal of Reliability}\label{sec:network}
Several recent studies have shown that both mainstream and alternative news sources often share (or copy) articles from each other either verbatim or in part~\cite{starbird2018ecosystem,horne2019different}. The motivation behind this content copying can differ greatly depending on the source. Mainstream sources copy articles from news-wire services often to meet demand or ``break'' news in a timely manner. Conspiracy sources may employ this tactic with malicious intent to spread false content, create uncertainty surrounding an event by amplifying alternative narratives, or to simply make money from clicks~\cite{horne2019different,starbird2018ecosystem,braun2019fake}. This behavior may also indicate coordination between disinformation producers. 

This article sharing behavior can be formulated as a network where each node is a news source and each directed edge $A\rightarrow B$ has weight proportional to the number of articles in $B$ that are copied from $A$. This network captures various important aspects of the news ecosystem: communities of similar media sources, hubs of conspiracy news production, and bridges between the mainstream and alternative media. It is likely that these network structures, particularly community membership, provide a strong signal of veracity. It is easy to imagine that an unknown news producer, which copies articles from a well-known conspiracy news producer, is also a source of conspiracy news. This signal can be extended to more indirect cases where unknown news sources fall in a path between two known news sources, or sources that copy from both reliable and unreliable sources can be labeled as mixed veracity. It is this rich structure of information that we wish to take advantage of in detecting articles from reliable and unreliable sources. 

\subsection{Network Construction}
Using the whole NELA-GT-2018 data set (rather than our extracted labeled data set described in Section~\ref{sec:data}), we follow the process described in \cite{horne2019different} to create a near-verbatim content sharing network (CSN) of news sources. Specifically, we compute a TF-IDF matrix of all articles in the data set and compute the cosine similarity between each article vector pair (given that each article comes from a different news source). To reduce the complexity of this process, we use a sliding 5 day window of articles. For each pair of article vectors that have a cosine similarity greater or equal to $0.85$, we extract them and order them by the timestamps. This is the same cosine similarity threshold used in both \cite{horne2019different} and \cite{starbird2018ecosystem}. This process creates a directed graph $G= (V, E)$, where $V$ is the set of news sources and $E$ are directed weighted edges representing articles shared. Edges are directed towards publishers that copied articles (inferred by the timestamps). We normalize the weight of each edge in the network by the number of articles published in total by the source. For example, if USA Today publishes $1000$ articles and copies $100$ of those articles from Reuters, the edge from USA Today to Reuters would have weight $0.1$ and be directed towards USA Today. 

We show a visualization of this constructed network in Figure~\ref{fig:nets1}. We built this visualization using Gephi and used the Newman Spectral Method for directed modularity to label community membership\cite{newman2004finding}. Specifically, we use the default parameters from Zhiya Zuo's modularity maximization Python package\footnote{\url{zhiyzuo.github.io/python-modularity-maximization/}}. This network includes the 52 sources with known labels used in this study as well as 88 sources with no labels. The presence of both labeled and unlabeled sources provides us with a rich network structure. In addition, the community structures in the network (as shown by colors in Figure~\ref{fig:nets1}) would likely be lost if only labeled data was used. We find that the community structure looks very similar to the structure displayed in \cite{horne2019different}, as we use the same dataset and the same community detection method. To provide some intuition of where labeled sources are placed in the network, we show the number of sources from each class in each network community in Figure~\ref{fig:members}. We also show the degree distributions of labeled sources in Figure~\ref{fig:dists}. Both Figures support the idea that \texttt{reliable} and \texttt{unreliable} sources are represented differently in the CSN, hence are potentially valuable in news veracity detection.

We choose to focus on near-verbatim content sharing networks in this paper due to the well-studied properties of these networks. Partial content sharing networks can also provide useful additional information, however these networks are not yet studied in the literature. Hence, we leave study of partial sharing behavior to future work.

\begin{table}[htbp]
  \centering
  \hspace*{-0.2in}\begin{tabular}{c}
   \includegraphics[width=3.0in]{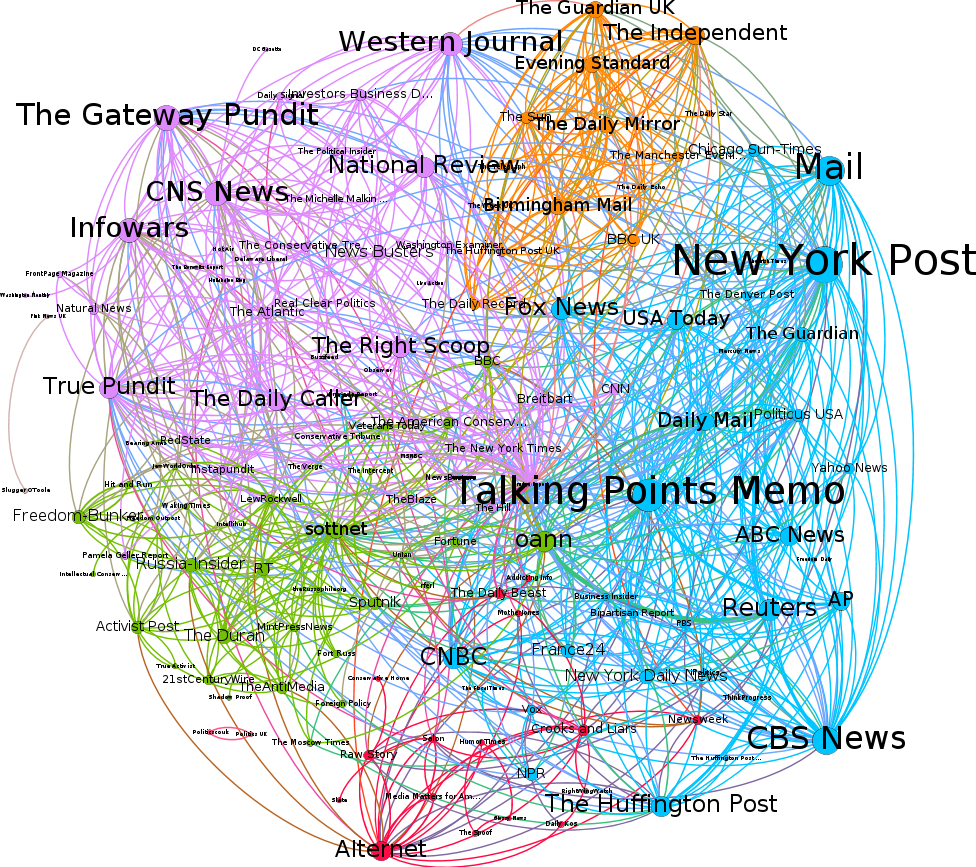}
  \end{tabular}
   \captionof{figure}{Visualization of CSN using Gephi~\cite{ICWSM09154}. Colors represent communities using directed modularity. Edges are directed, where the outdegree of node $n$ is how many news sources copy articles from node $n$. The size of each node is based on outdegree. Just as shown in \cite{horne2019different}, each community contains sources from distant parts of the media landscape, often grouping sources on similar veracity. In particular, we can see many of our unreliable sources in the magenta and green communities, while our reliable sources fall mostly within the blue community.}
  \label{fig:nets1}
\end{table}

\begin{table}[htbp]
  \centering
  \hspace*{-0.0in}\begin{tabular}{c}
   \includegraphics[width=2.5in]{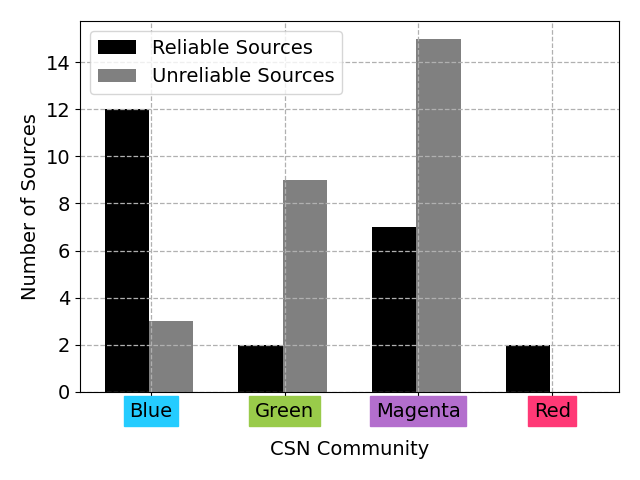}
  \end{tabular}
   \captionof{figure}{Number of labeled sources in each community, where the colors of the x axis labels correspond to communities in Figure~\ref{fig:nets1}. The high separation between \texttt{reliable} and \texttt{unreliable} labeled sources supports the intuition that the CSN network can be used to approximate veracity.}
  \label{fig:members}
\end{table}


\subsection{Network Representation for Classification}
\textbf{Hand-crafted Network Features (HCNF).} One way to represent sources in the CSN is to craft a set of network features for each source. To do this, we choose several standard network measures, as well as more community-focused features. In total we compute 11 features that include:
\begin{enumerate}
    \item Community - What community is the source in, as determined by directed modularity \cite{newman2004finding}.
    \item Weighted Outdegree - Number of articles copied from the source.
    \item Weighted Indegree - Number of articles the source copies.
    \item Closeness Centrality - The reciprocal of the sum of the shortest path distances from the source to all other sources.
    \item Betweenness Centrality - The sum of the fraction of all-pairs shortest paths that pass through the source.
    \item Eigenvector Centrality - The centrality of a source based on the centrality of its neighboring sources.
    \item Community Core - Is the source a member of the k-core of its community or not, where the k-core is a maximal subgraph that contains nodes of degree k or more. We compute the the core with the largest degree.
    \item Inside Source Edges - Number of articles copied from the source by sources in its community.
    \item Inside Sink Edges - Number of articles the source copies from other sources in its community.
    \item Importing Edges - Number of articles the source copies from sources outside of its community.
    \item Exporting Edges - Number of articles copied by the source from sources outside of its community.
\end{enumerate}

\begin{table}[htbp]
  \centering
  \hspace*{-0.0in}\begin{tabular}{cc}
   \includegraphics[width=0.22\textwidth]{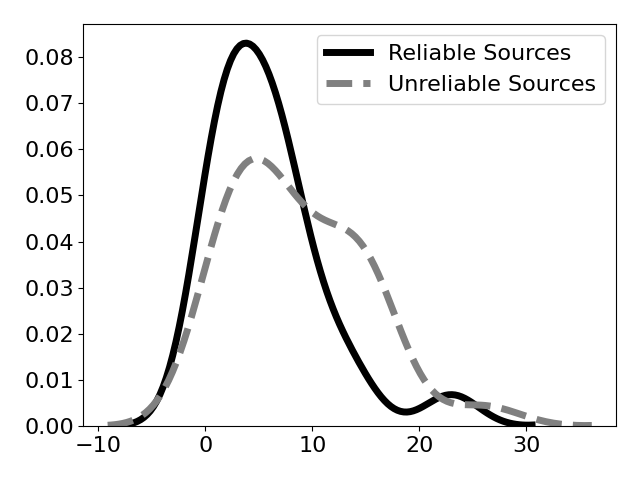}
   &\includegraphics[width=0.22\textwidth]{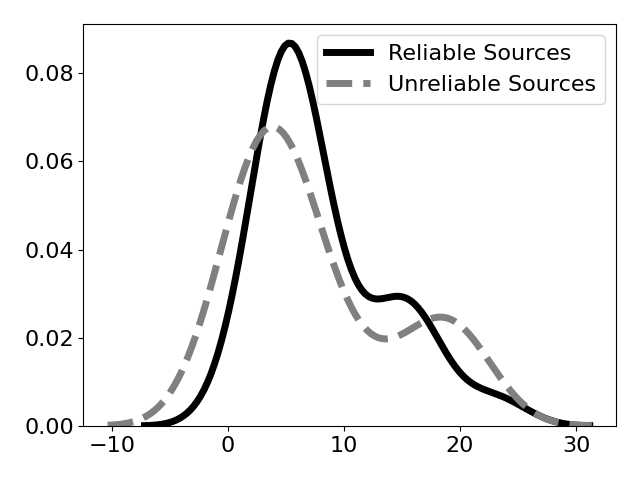}\\
   (a) In-Degree Distribution & (b) Out-Degree Distribution
  \end{tabular}
   \captionof{figure}{Degree distributions \texttt{reliable} and \texttt{unreliable} labeled sources in CSN, where In-Degree (a) represents how much a source is copying articles and (b) Out-Degree represents how much a source is copied from. As discussed in literature~\cite{horne2019different}, unreliable sources generally copy more articles verbatim than reliable sources.}
  \label{fig:dists}
\end{table}

\textbf{\textbf{Node2Vec (N2V)}.}
Another, likely more complete, method to represent the CSN network is network embedding. Specifically, we use the Node2Vec \cite{grover2016node2vec} network embedding method. As with word embedding, Node2Vec uses the skipgram model and transforms the sparse adjacency matrices of networks into a dense vector representation of nodes. This representation aims to preserve network structure and node neighborhood, clustering together those nodes with similar functionality and structure in the network, such as hubs and peripheral nodes. Additionally, the dense vector representation captures latent similarity relations within the network. Node2Vec uses the return parameter $p$ and the in-out parameter $q$ to control the breadth and depth of random walks on the network used to generate the embedding. In this work, we use $p=0.5, q=0.5$ and set the vector dimension to $40$. The output are vectors representing the network nodes (news sources), we refer to these vectors as N2V features. Note that we embed all sources in our dataset, including those with no labels to fully represent the CSN. Note, we remove all articles used in the CSN construction from our training and test data in later experiments in order to avoid data leakage.

\textbf{NetworkText2Vec (NT2V).}
Naturally, the CSN embedding can only represent sources that share content. The sharing behavior may be rare and not present in specific settings. Furthermore, not all sources may share content in verbatim, especially if they are new or are representing different topics. For example, a source may focus on breaking news, which lends itself to content sharing, while another source may focus on investigative pieces, which may not lend itself to content sharing. Although these sources can be represented as completely disconnected nodes in the network, embedding disconnected nodes with N2V would give us no relevant information with respect to node similarity. To fill this gap, we can use the similarity of sources with respect to the text that they publish, using text as side information in the embedding. 

The problem of attributed network embedding was addressed by Yang et al. \cite{yang2015network}, they proposed TADW, a method that performs matrix factorization on a matrix using as input both network and text features. One limitation of this method is that it requires network and text representation of every source being embedded. In our case, this is a major drawback, as CSNs can have missing edges due to lack of content sharing information. To mitigate this issue, we propose a method based the multi-scale attributed network embedding by \cite{rozemberczki2019multi} that we refer to as \OUREMBEDDING. This method takes as input both the CSN and the text attributes of news articles. The text attributes are a representation of a source given by the average of its word embedding vectors. Using this information, \OUREMBEDDING~ combines two random walks based on the similarity of nodes (news sources): one over the network as in Node2Vec and the second one over the text attributes. More formally, let $c_i$ be the $i$-th source from a random walk over the text corpus, the transition probabilities from $c_{i}$ to $c_{i+1}$ are obtained from the cosine similarity between $c_{i}$ and its $k$-nearest neighbors, and normalized by the sum of weights of the edges leaving $c_i$. Sources with higher cosine similarities have higher chances of being picked in the random walk, thus appearing more often in contexts. We set a lower bound cutoff similarity of $0.5$ to prevent selecting sources that are significantly dissimilar. 

Intuitively, the process of context generation is carried out by interchanging random walks over the network space and the text space. At random walk $j$ we decide with probability $t$ that the walk will happen over the text space, or $1-t$ that it will happen over the network. If the network is chosen, we perform a random walk entirely over the network, as with Node2Vec, otherwise the random walk is entirely over the text corpus space. We generate $n$ contexts for each source. Once contexts are generated, they are used as the input to a skipgram model. In addition to the input parameters $p$ and $q$ of Node2Vec, \OUREMBEDDING~requires the $t$ parameter that controls the likelihood of performing a walk over the text and the $k$ parameter that controls the number of nearest neighbors to consider during the text corpus walk. The output is vector representations of sources based on the generated contexts. 

We set the output vector size to $40$, number of walks to $1000$, walk length to 80, and tune parameters $p$, $q$ and $t$ by performing a grid search over the interval $[0.2,0.8]$ with a step size of $0.1$. We select the model that yielded the best classification accuracy on a validation set. The final parameters are: $t=0.8$, $p=0.5$, $q=0.4$. Code for NT2V and additional documentation are publicly available\footnote{\url{https://github.com/mgruppi/NewsNetworkEmbedding}}. We uniformly sample 20\% of the articles for each source to use in NT2V, the sampled articles are used exclusively to compute the source representation and are not used in any other scenario. This is done in order to avoid data leakage from the source representation into the article level experiments.

\section{Baseline Text Models}
To compare our CSN feature models to state-of-the-art text-based methods, we compute several text feature sets and discuss the details of each below. 

\textbf{NELA.}
NELA is a hand-crafted, text feature set used in whole or in part in several news veracity studies~\cite{horne2018accessing,horne2019robust,baly2018,baly2019multi,cruz2019team,barron2019proppy} with available code online\footnote{\url{https://pypi.org/project/nela-features/}}. This feature set can be divided into five different groups: 
\begin{enumerate}
    \item Style - This group represents the general writing style of an article, including parts-of-speech used, punctuation used, and capitalization used. 
    \item Complexity - This group measures the complexity of writing in an article, including lexical diversity, reading difficulty, length of word, and length of sentences. 
    \item Bias - This group measures the bias of an article. It uses lexicons developed in \cite{recasens2013linguistic} to capture various signals of bias in text, such as hedges, factives, assertives, and opinion words.
    \item Affect - This group measures the emotion and sentiment of text using two well-known works in text processing: LIWC~\cite{tausczik2010psychological} and VADER~\cite{hutto2014vader}. LIWC is a gold-standard, lexicon based method for discovering various social and psychological traits in text. These include various types of emotion, such as anger, anxiety, affect, and swear words. VADER is a state-of-the-art sentiment detection tool that provides measures of positive, negative, and neutral emotion in text. 
    \item Moral - This feature group is a lexicon based method that measures morality in text on the basis of Moral Foundation Theory~\cite{graham2013moral}. Examples of these features include fairness, authority, and care. 
\end{enumerate}

In total NELA contains 194 features, computed independently on the body text and title text of an article. 

\textbf{FastText (FT).}
Another method we can use to capture textual differences between news articles is using word embedding. Word embedding features have been used in only a few news veracity detection studies so far~\cite{singhania20173han} and are still under-explored. The potential advantage of word embedding features over hand-crafted feature sets, like NELA, is that features can be automatically captured regardless of language and domain. The disadvantage is that we cannot control the specific concepts captured in the text, which may lead to worse performance and robustness.

In this work, we use the \texttt{wiki-news-300d-1M}\footnote{\url{https://fasttext.cc/docs/en/english-vectors.html}} pre-trained FastText model ~\cite{bojanowski2017enriching} to obtaine the representation for \NUMARTICLES news articles, the model was pre-trained on Wikipedia and news data, contains 1 million words and the vector dimension is 300. To obtain the representation of an entire news article we average the vectors of all the words in an article's title and content, thus, arriving at the final representation of an article, given by a $300$ dimension article vector which we refer to as FT features.

Note, we also experiment with a LSTM sequence classifier and BERT embedding vectors as baseline text models, but due to the similarity of results across the text models and space restrictions, we do not display those results. 

\begin{table*}[]
\fontsize{9.5pt}{9.5pt}
\begin{tabular}{clcccc}
    \multicolumn{1}{c}{} & \textbf{Feature Group} & \textbf{Accuracy}  &   \textbf{F1 Score} & \textbf{Precision} & \textbf{Recall}  \\ \specialrule{2pt}{1pt}{1pt}
    \multirow{3}{*}{\textbf{TX only}}   &   \textbf{NELA}               &   0.689   &   0.685 & 0.692  & 0.741   \\
                                        &   \textbf{FT}                 &   0.636   &   0.613 & 0.682  & 0.637   \\
                                        &   \textbf{FT+NELA}            &   0.681   &   0.685 & 0.700  & 0.719  \\ \hline
    \multirow{2}{*}{\textbf{NT only}}   &   \textbf{HCNF}               &   0.778   &   0.773 & 0.827  & 0.767  \\
                                        &   \textbf{N2V}                &   0.802   &   0.813 & \textbf{0.860}  & 0.815  \\ \hline
    \multirow{9}{*}{\textbf{TX + NT}}  &   \textbf{NELA+HCNF}           &   0.733   &   0.723 & 0.768  & 0.736  \\
                                        &   \textbf{FT+HCNF}            &   0.730   &   0.707 & 0.739 & 0.735  \\
                                        &   \textbf{NELA+N2V}           &   0.789   &   0.791 & 0.808 & 0.789  \\
                                        &   \textbf{FT+N2V}             &   \textbf{0.836}   &   0.820 & 0.841 & 0.842  \\
                                        &   \textbf{NT2V}               &   0.802   &   0.803 & 0.790 & 0.860  \\
                                        &   \textbf{NELA+NT2V}          &   0.788   &   0.773 & 0.802  & 0.835  \\
                                        &   \textbf{FT+NT2V}            &   0.805   &   0.806 & 0.824  & 0.834  \\
                                        &   \textbf{Ensemble between NT2V \& NELA}      &   0.817   &   \textbf{0.823} & 0.798 & \textbf{0.893}  \\
                                        &   \textbf{Ensemble between NT2V \& FT}        &   0.806   &   0.793 & 0.798 & 0.840  \\ \specialrule{2pt}{1pt}{1pt}
    
\end{tabular}
\caption{Average performance scores over 50 runs of 20\% sources as a test set. \OUREMBEDDING params: $t=0.8$ $p=0.5$ $q=0.4$. The best scores in each category are shown in bold font. See Section~\ref{result1} for details on $+$ and ensemble combinations.}
\label{tbl:scores}
\end{table*}

\section{Results}
\subsection{CSN features improve the accuracy of text-based models}\label{result1}
Again, the goal of our classification model is to predict if the source of a news article is reliable or unreliable, given a news article and its source name as input. To this end, we train Random Forest classifiers on 80\% of the sources and test on 20\% of the sources. For each source, we uniformly sample 1000 articles before splitting into train and test sets to ensure each test set is balanced. Note, we are simulating a setting in which the classifier is given an individual news article from an unknown source as input and uses both article-related features and source-related features to predict. If a source is selected for testing, all 1000 of its sampled articles are removed from training. We repeat this experiment 50 times and average the performance metrics. We also repeat these experiments using a fully-connected Neural Network classifier, but find little to no improvement over the Random Forest Classifier, hence we only display the results using Random Forest.

To assess how much CSN features and text features contribute to distinguishing articles from reliable and unreliable sources, we test each individual feature group as well as combinations of article-level text features with their respective source-level CSN features. We combine text and CSN features in two ways:

\begin{enumerate}
    \item We concatenate text and CSN vectors (represented with a plus sign, e.g. NELA+N2V) and predict using a single binary classifier, or
    \item We use a feature ensemble of two binary classifiers, one trained on text features and the other trained on CSN features, using the sigmoid function to predict a probability that the given input belongs to class $0$ (reliable). Those probabilities are then combined using a soft voting.
\end{enumerate}

Table \ref{tbl:scores} shows the classification results for all feature group combinations and classification algorithms. As shown in Table \ref{tbl:scores}, both the hand-crafted text model (NELA) and the word embedding model (FT) are improved by the CSN features (N2V and NT2V). These improvements are significant, increasing accuracy as much as 20\%. Based on overall accuracy, the best model is FT+N2V, while the feature Ensemble using NELA shows the best F1 and Recall scores. 

While the best performing models are all using combinations of the CSN features and the text features, we do see the CSN models alone also perform well. In fact, N2V has the best precision score among all models and has only a 3\% decrease in accuracy from the best combination model, demonstrating the strong signal provided by the CSN.

\subsection{Text models and CSN models often make different mistakes}
It is clear that CSN features capture some signal of veracity and improve upon the text-based models. However, do CSN models make the same mistakes as the traditionally used text models? To test this, we use two methods. First, we compute the conditional probabilities that a feature correctly classifies the articles given that another feature set has failed to classify it, shown in Table~\ref{tbl:leaveoneout}. More precisely, given feature sets $A$ and $B$, we compute $P(p_B=1|p_A=0)$ as the \emph{conditional accuracy}, where $p_B=1$ is the event where feature set B correctly classifies an article, and $p_A=0$ is the event where feature set A does not correctly classify the same article. The probabilities were computed using a classification model trained on a \emph{leave one source out} subset of articles. Specifically, for each source $s$, let $\mathcal{S}$ be the articles from $s$ in the data $\mathcal{D}$. We train a Random Forest classifier on $\mathcal{D} - \mathcal{S}$, and test the classification on $\mathcal{S}$. The conditional probability indicates how many of the mistakes of $A$ are corrected by $B$, and it is given by $P(p_B=1|p_A=0) = \frac{P(p_B=1) \cap P(p_A=0)}{P(p_A=0)}$.

Second, we examine the distribution of errors per class for each feature group, shown in Table~\ref{tbl:err}. Simply put, using the \emph{leave one source out} method, we calculate what proportion of the wrong classifications are in each class.  This analysis shows us which feature groups are better or worse at classifying one class or the other. 

\begin{table}[!htb]
\centering
\fontsize{9.5pt}{9.5pt}
\begin{tabular}{cccc}
\multicolumn{1}{c}{\textbf{$A$}} & \multicolumn{1}{c}{\textbf{$B$}} &
\multicolumn{1}{c}{\textbf{$P(p_A=0)$}} & 
\multicolumn{1}{c}{\textbf{$P(p_B=1|p_A=0)$}} \\ \specialrule{2pt}{1pt}{1pt}
\textbf{FT}                    & \textbf{N2V}   & 0.38                   & 0.79                                    \\
\textbf{N2V}                   & \textbf{FT}    & 0.17                   & 0.54                                    \\\hline
\textbf{FT}                    & \textbf{NT2V}  & 0.38                  & 0.72                                    \\
\textbf{NT2V}                  & \textbf{FT}    & 0.23                    & 0.35                                    \\\hline
\textbf{NELA}                  & \textbf{N2V}   & 0.33                   & 0.83                                    \\
\textbf{N2V}                   & \textbf{NELA}  & 0.17                  & 0.66                                    \\\hline
\textbf{NELA}                  & \textbf{NT2V}  & 0.33                  & 0.73                                    \\
\textbf{NT2V}                  & \textbf{NELA}  & 0.23                  & 0.63                                    \\\hline
\textbf{FT}                    & \textbf{NELA}  & 0.38                  & 0.43                                    \\
\textbf{NELA}                  & \textbf{FT}    & 0.33                  & 0.33
                    \\\hline
\textbf{FT}                     & \textbf{HCNF} & 0.38                  & 0.75                                  \\
\textbf{HCNF}                   & \textbf{FT}   & 0.18                  & 0.45                                  \\\hline
\textbf{NELA}                   & \textbf{HCNF} & 0.33                  & 0.75                                  \\
\textbf{HCNF}                   & \textbf{NELA} & 0.18                  & 0.54                                  \\\hline
\textbf{N2V}                    & \textbf{HCNF} & 0.16                  & 0.61                                  \\ 
\textbf{HCNF}                   & \textbf{N2V}  & 0.18                  & 0.66                              
                            \\
\specialrule{2pt}{1pt}{1pt}\\

\end{tabular}
\caption{Conditional probabilities of mistakes made by each feature set. $P(p_A=0)$ is the probability of feature set $A$ making a mistake, $P(p_B=1|p_A=0)$ is the conditional accuracy defined as the probability that feature set B correctly classifies samples given that feature set $A$ failed to do so. The higher the probability, the more dissimilar the mistakes made by each feature set is. Each model uses Random Forest. We use bold font to indicate the highest dissimilarity between CSN models and text models and vice versa.}
\label{tbl:leaveoneout}
\end{table}

\begin{table}[]
    \centering
    \fontsize{9.5pt}{9.5pt}
    \begin{tabular}{lcc}
        & \multicolumn{2}{c}{\textbf{Error rate}}  \\
  
        \multicolumn{1}{l}{ \textbf{Feature Group}}                 & \textbf{Reliable}    & \textbf{Unreliable} \\         \specialrule{2pt}{1pt}{1pt}   
        \textbf{NELA}    & 0.32                 & 0.68  \\
        \textbf{FT}      & 0.14                 & 0.86  \\
        \textbf{HCNF}    & 0.41                 & 0.59  \\
        \textbf{N2V}     & 0.63                 & 0.37  \\
        \textbf{NT2V}    & 0.81                 & 0.19  \\
        \specialrule{2pt}{1pt}{1pt} 
    \end{tabular}
    \caption{Distribution of errors per class per feature group using a Random Forest classifier. Content based models show better performance when classifying reliable sources, whereas CSN features shows better performance when classifying unreliable sources.}
    \label{tbl:err}
\end{table}


As shown in Table~\ref{tbl:leaveoneout}, the CSN models (HCNF, N2V, NT2V) made very different mistakes than the text models (NELA, FT), with at most a 83\% chance of a CSN model correctly classifying an article that a text model missed. When reversing the probability, we similarly see different mistakes made, with at most a 66\% chance of a text model correctly classifying an article that a CSN model missed. When looking at the specific types of mistakes made, we see several consistent cases. Generally, we see the same trend in Table~\ref{tbl:err}. Specifically, we see that both NELA and FT (text models) are better at classifying the \texttt{reliable} class than the \texttt{unreliable} class, while N2V and NT2V (network models) are much better at classifying the \texttt{unreliable} class than the \texttt{reliable} class. Higher conditional probabilities imply greater distinction between the errors made by one feature group and the other.

Overall, there are very few mistakes by the CSN features, but when they do make mistakes, it is on sources in sparsely labeled areas of the network. For example, in our data set, Reuters and The Guardian, are often mis-classified by N2V (i.e. purely CSN information). This mistake is because articles from both Reuters and The Guardian are often copied by U.K mainstream sources, which are unlabeled in our data set. Thus, when Reuters or The Guardian are left-out for testing, the model has very few, if any, examples of reliable sources in the same U.K neighborhood. However, when text is added to the model in some way, whether that is through the text space in NT2V sampling or the purely text-based features, these mistakes are corrected, as there are numerous examples of reliable sources that produce articles similar to the style of Reuters and The Guardian. In this specific case, the CSN model can be improved by having more labels in the non-U.S. communities of the network. This sparse label problem is also why we see the CSN models classifying the \texttt{unreliable} class better, as the unreliable sources are more densely clustered together than the reliable sources in the network. We leave explicit tests on the impact of removing and adding nodes/labels in the CSN to future work.

Another interesting case is when both the CSN and text features incorrectly label an article, but the combination of them flips the label. For example, some articles from Business Insider, a reliable news source, are classified in this way. In the CSN space, Business Insider falls in the U.S. mainstream community, but is a peripheral node, which may lead to very few other reliable nodes being sampled in the network embedding process. In the text space, the articles are similar to other mainstream sources in the body, but the titles can sometimes be considered `clickbait', which is often a trait of unreliable news articles. Hence, both feature models individually may not have enough information to say it is similar to a reliable source, but together they can correctly label the article. 

We also note that not all text models are alike. We found that the hand-crafted text features (NELA) and the word representation features (FT) also make dissimilar mistakes. While these mistakes are not as dissimilar as those between the text-based models and the CSN models, they are notably different, with a 33\% chance that a mistake made by NELA is correctly classified by FT, and a 43\% vice versa. However, these differences in mistakes do not seem to be enough to help prediction performance. When qualitatively looking at these differences in mistakes, it is hard to say what specifically causes them. However, in general, it seems they are capturing different features of the text. A clear example of these differences is in the titles of articles from unreliable sources. FT often mis-classifies articles from unreliable sources with the word `BREAKING' in the headline or with `clickbait' headlines. NELA correctly classifies these articles. This result makes sense, as NELA has many features which focus specifically on the title of articles, including the use of words in all capital letters, while a word embedding model cannot directly capture this.


\begin{table}[]
    \centering
    \fontsize{9.5pt}{9.5pt}
    \begin{tabular}{lcc}
        \textbf{Feature Group} & \textbf{In Time} & \textbf{Forecast} \\        \specialrule{2pt}{1pt}{1pt}   
        \textbf{FT} & 0.61 & 0.55 \\
        \textbf{FT+NT2V} & 0.75 & 0.72  \\
        \textbf{NELA} & 0.63 & 0.59   \\
        \textbf{NELA+NT2V} & 0.71 & 0.64  \\
        \textbf{NELA+HCNF} & 0.66 & 0.69  \\
        \textbf{FT+HCNF} & 0.55 & 0.60   \\
        \specialrule{2pt}{1pt}{1pt}   
    \end{tabular}
    \caption{Accuracy scores for each feature group when testing In Time (first month), and when forecasting (testing on a time period outside of training). The results show that the combination of text and network (NT2V) features  has better forecast potential then the text-only models. A Random Forest classifier was used.}
    \label{tab:forecast}
\end{table}

\begin{figure}[ht!]
    \begin{subfigure}{0.45\textwidth}
        \centering
        \includegraphics[width=\textwidth]{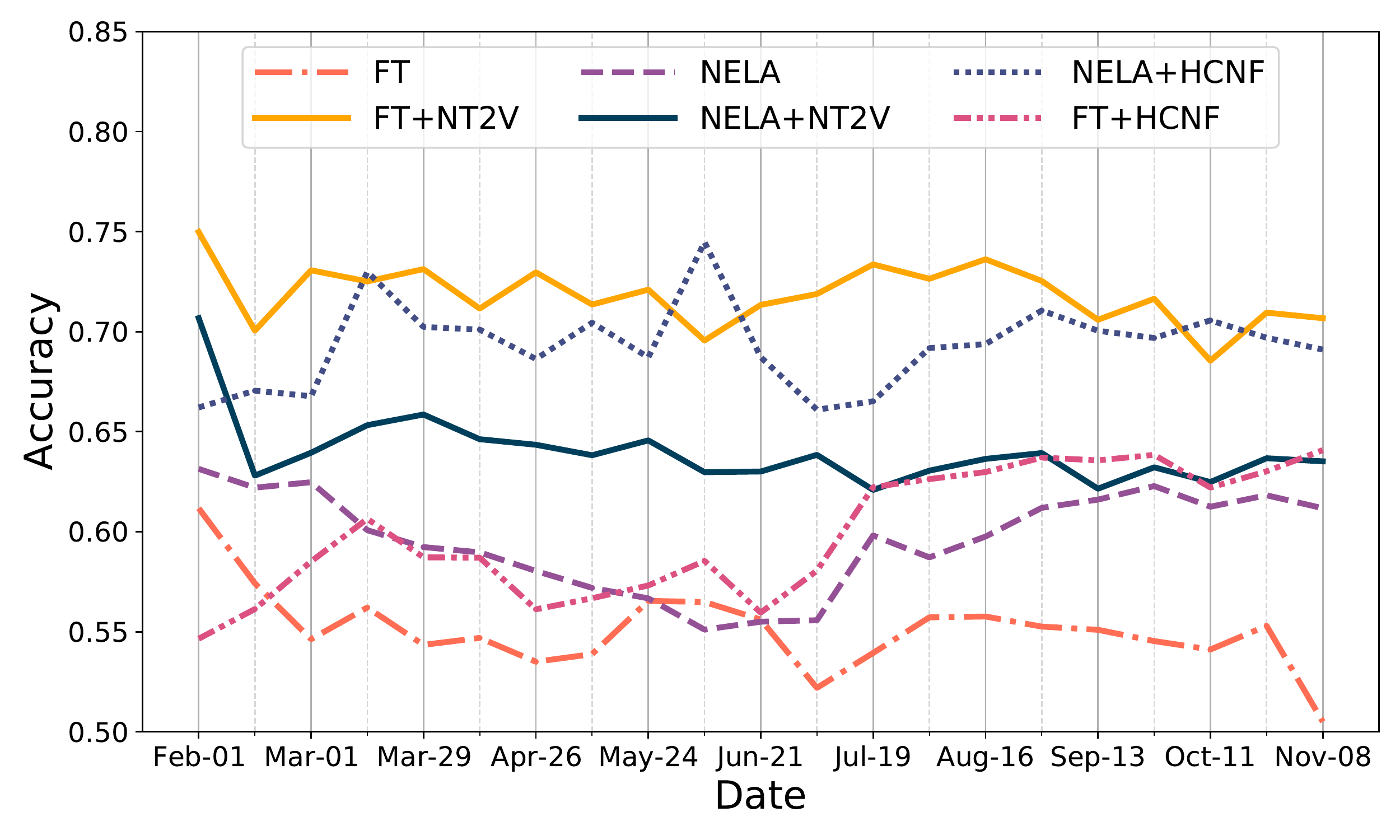}
    \end{subfigure}
    \caption{Classification accuracy overtime. The first month of data is used for training the classifiers, which are tested on each subsequent two week time slices. The combination of text and CSN features provide higher accuracy and stability over time, particularly with the combination of FT and NT2V features.}
    \label{fig:overtime_plot}
\end{figure}

\subsection{CSN features improve the stability of text-based models over time}
In this section, we examine the stability of the performance improvements from network models over time. To test this, we train each classifier on the first month of data and test the classifier on each 2 week slice of data moving forward in time. We only test the models on sources that are unused in training and perform this train-test split over 50 runs of 20\% of the sources. Again, we ensure that each source is balanced. Note, we also reconstruct the CSN network to only include information from the first month of data. This simulates a classifier that is built in February 2018 and left static for the rest of the year. These results are in Figure~\ref{fig:overtime_plot}. 

In addition to showing performance stability over time for each model, in Table \ref{tab:forecast}, we show the classification accuracy for each feature group in two scenarios, using a Random Forest classifier: \texttt{in time} and \texttt{forecast}. The \texttt{in time} test is a prediction test on data from the same time period as the training (i.e. February 2018), while the \texttt{forecast} test is a prediction test on the remaining time period without re-training.

As shown in Figure~\ref{fig:overtime_plot}, the addition of NT2V features improves both the overall performance of the model and its consistency over time. For example, for FT, there is at most an accuracy drop of 11\% over 10 months (0.61 to 0.50). However, when combined with NT2V, not only the initial accuracy is higher, but the drop is more subtle (0.75 to 0.70). However, not all NT2V combinations remain this stable. Specifically, when NT2V is combined with NELA, we similarly see a boost in overall accuracy, but see a significant initial drop in accuracy from February to March (-8\%). However, the model remains very stable after the initial drop.

The results in Table \ref{tab:forecast} show that combining text and network features improve the forecast performance, but this performance increase is not always significant.

\section{Discussion and Conclusion}
In this study, we presented a novel feature set for the detection of articles from unreliable sources, utilizing the rich structure of news content sharing networks. To do this, we used a network embedding method that takes a deep walk approach to sample from both the CSN space and the text space. The addition of the text space to the CSN space in the sampling process makes it possible to find representations of incomplete networks by positioning sources with unknown CSN information close to those with high similarity in the text space. We show that the information provided by embedding CSN networks provides a strong signal of reliability and boosts the accuracy of text-based models. We show that text information and CSN information make dissimilar mistakes, illustrating complementary signals between the two types of models. Saliently, these CSN features also stabilize the performance of text-based models over time, performing consistently over a 10 month time frame without retraining. This stabilization is likely due to the fact that the CSN structure remains largely unchanged over time, while text features are vulnerable to recurrent topic changes. 

There may be additional advantages to using the CSN embedding model that can be explored in future work. First, both the CSN construction and word embedding are language-agnostic, unlike the hand-crafted text features (NELA). Assuming reliable and unreliable media operate distinctly in other languages and cultures, the NT2V embedding can be used out-of-the-box to detect these differences. In fact, this method could be extended to many other types of information spaces beyond political news, as it is common for sources to amplify their message by creating copies (e.g. bot-generated retweets on Twitter). Second, CSN features may also work in distinguishing different granularity of labels, due to the tightly-formed communities in the network. For example, if we have labels of political-leaning or other characteristics of sources, it is possible that we can separate them in the network space. 


In conclusion, using the behavior of information producers provides valuable signal in news veracity classification. This result points to a bigger picture need to explore and understand tactics used by disinformation producers not only for social interventions, but for automated support tools. If we can continue to structure information producer behaviors and tactics clearly, they can be used to aid our automated methods, which in turn can further our understanding of the news ecosystem. 



\bibliographystyle{ACM-Reference-Format}
\bibliography{references}


\begin{thebibliography}{46}


\ifx \showCODEN    \undefined \def \showCODEN     #1{\unskip}     \fi
\ifx \showDOI      \undefined \def \showDOI       #1{#1}\fi
\ifx \showISBNx    \undefined \def \showISBNx     #1{\unskip}     \fi
\ifx \showISBNxiii \undefined \def \showISBNxiii  #1{\unskip}     \fi
\ifx \showISSN     \undefined \def \showISSN      #1{\unskip}     \fi
\ifx \showLCCN     \undefined \def \showLCCN      #1{\unskip}     \fi
\ifx \shownote     \undefined \def \shownote      #1{#1}          \fi
\ifx \showarticletitle \undefined \def \showarticletitle #1{#1}   \fi
\ifx \showURL      \undefined \def \showURL       {\relax}        \fi
\providecommand\bibfield[2]{#2}
\providecommand\bibinfo[2]{#2}
\providecommand\natexlab[1]{#1}
\providecommand\showeprint[2][]{arXiv:#2}

\bibitem[\protect\citeauthoryear{Agerri}{Agerri}{2019}]%
        {agerri2019doris}
\bibfield{author}{\bibinfo{person}{Rodrigo Agerri}.}
  \bibinfo{year}{2019}\natexlab{}.
\newblock \showarticletitle{Doris Martin at SemEval-2019 Task 4: Hyperpartisan
  News Detection with Generic Semi-supervised Features}. In
  \bibinfo{booktitle}{\emph{Proceedings of the 13th SemEval}}.
  \bibinfo{pages}{944--948}.
\newblock


\bibitem[\protect\citeauthoryear{Allcott and Gentzkow}{Allcott and
  Gentzkow}{2017}]%
        {allcott2017social}
\bibfield{author}{\bibinfo{person}{Hunt Allcott} {and} \bibinfo{person}{Matthew
  Gentzkow}.} \bibinfo{year}{2017}\natexlab{}.
\newblock \showarticletitle{Social media and fake news in the 2016 election}.
\newblock \bibinfo{journal}{\emph{Journal of Economic Perspectives}}
  \bibinfo{volume}{31}, \bibinfo{number}{2} (\bibinfo{year}{2017}),
  \bibinfo{pages}{211--36}.
\newblock


\bibitem[\protect\citeauthoryear{Alvermann}{Alvermann}{2017}]%
        {alvermann2017social}
\bibfield{author}{\bibinfo{person}{Donna~E Alvermann}.}
  \bibinfo{year}{2017}\natexlab{}.
\newblock \showarticletitle{Social Media Texts and Critical Inquiry in a
  Post-Factual Era}.
\newblock \bibinfo{journal}{\emph{Journal of Adolescent \& Adult Literacy}}
  \bibinfo{volume}{61}, \bibinfo{number}{3} (\bibinfo{year}{2017}),
  \bibinfo{pages}{335--338}.
\newblock


\bibitem[\protect\citeauthoryear{Baly, Karadzhov, Alexandrov, Glass, and
  Nakov}{Baly et~al\mbox{.}}{2018}]%
        {baly2018}
\bibfield{author}{\bibinfo{person}{Ramy Baly}, \bibinfo{person}{Georgi
  Karadzhov}, \bibinfo{person}{Dimitar Alexandrov}, \bibinfo{person}{James
  Glass}, {and} \bibinfo{person}{Preslav Nakov}.}
  \bibinfo{year}{2018}\natexlab{}.
\newblock \showarticletitle{Predicting Factuality of Reporting and Bias of News
  Media Sources}. In \bibinfo{booktitle}{\emph{Proceedings of 2018 EMNLP}}.
\newblock


\bibitem[\protect\citeauthoryear{Baly, Karadzhov, Saleh, Glass, and Nakov}{Baly
  et~al\mbox{.}}{2019}]%
        {baly2019multi}
\bibfield{author}{\bibinfo{person}{Ramy Baly}, \bibinfo{person}{Georgi
  Karadzhov}, \bibinfo{person}{Abdelrhman Saleh}, \bibinfo{person}{James
  Glass}, {and} \bibinfo{person}{Preslav Nakov}.}
  \bibinfo{year}{2019}\natexlab{}.
\newblock \showarticletitle{Multi-Task Ordinal Regression for Jointly
  Predicting the Trustworthiness and the Leading Political Ideology of News
  Media}.
\newblock \bibinfo{journal}{\emph{arXiv preprint arXiv:1904.00542}}
  (\bibinfo{year}{2019}).
\newblock


\bibitem[\protect\citeauthoryear{Barr{\'o}n-Cedeno, Jaradat, Da~San~Martino,
  and Nakov}{Barr{\'o}n-Cedeno et~al\mbox{.}}{2019}]%
        {barron2019proppy}
\bibfield{author}{\bibinfo{person}{Alberto Barr{\'o}n-Cedeno},
  \bibinfo{person}{Israa Jaradat}, \bibinfo{person}{Giovanni Da~San~Martino},
  {and} \bibinfo{person}{Preslav Nakov}.} \bibinfo{year}{2019}\natexlab{}.
\newblock \showarticletitle{Proppy: Organizing the news based on their
  propagandistic content}.
\newblock \bibinfo{journal}{\emph{Info. Proc. \& Manage.}}
  (\bibinfo{year}{2019}).
\newblock


\bibitem[\protect\citeauthoryear{Bastian, Heymann, and Jacomy}{Bastian
  et~al\mbox{.}}{2009}]%
        {ICWSM09154}
\bibfield{author}{\bibinfo{person}{Mathieu Bastian}, \bibinfo{person}{Sebastien
  Heymann}, {and} \bibinfo{person}{Mathieu Jacomy}.}
  \bibinfo{year}{2009}\natexlab{}.
\newblock \bibinfo{title}{Gephi: An Open Source Software for Exploring and
  Manipulating Networks}.
\newblock
\newblock
\urldef\tempurl%
\url{http://www.aaai.org/ocs/index.php/ICWSM/09/paper/view/154}
\showURL{%
\tempurl}


\bibitem[\protect\citeauthoryear{Boczkowski}{Boczkowski}{2010}]%
        {boczkowski2010news}
\bibfield{author}{\bibinfo{person}{Pablo~J Boczkowski}.}
  \bibinfo{year}{2010}\natexlab{}.
\newblock \bibinfo{booktitle}{\emph{News at work: Imitation in an age of
  information abundance}}.
\newblock \bibinfo{publisher}{University of Chicago Press}.
\newblock


\bibitem[\protect\citeauthoryear{Bojanowski, Grave, Joulin, and
  Mikolov}{Bojanowski et~al\mbox{.}}{2017}]%
        {bojanowski2017enriching}
\bibfield{author}{\bibinfo{person}{Piotr Bojanowski}, \bibinfo{person}{Edouard
  Grave}, \bibinfo{person}{Armand Joulin}, {and} \bibinfo{person}{Tomas
  Mikolov}.} \bibinfo{year}{2017}\natexlab{}.
\newblock \showarticletitle{Enriching word vectors with subword information}.
\newblock \bibinfo{journal}{\emph{Transactions of the ACL}}
  \bibinfo{volume}{5} (\bibinfo{year}{2017}), \bibinfo{pages}{135--146}.
\newblock


\bibitem[\protect\citeauthoryear{Brady, Wills, Jost, Tucker, and
  Van~Bavel}{Brady et~al\mbox{.}}{2017}]%
        {brady2017emotion}
\bibfield{author}{\bibinfo{person}{William~J Brady}, \bibinfo{person}{Julian~A
  Wills}, \bibinfo{person}{John~T Jost}, \bibinfo{person}{Joshua~A Tucker},
  {and} \bibinfo{person}{Jay~J Van~Bavel}.} \bibinfo{year}{2017}\natexlab{}.
\newblock \showarticletitle{Emotion shapes the diffusion of moralized content
  in social networks}.
\newblock \bibinfo{journal}{\emph{Proceedings of the National Academy of
  Sciences}} \bibinfo{volume}{114}, \bibinfo{number}{28}
  (\bibinfo{year}{2017}), \bibinfo{pages}{7313--7318}.
\newblock


\bibitem[\protect\citeauthoryear{Braun and Eklund}{Braun and Eklund}{2019}]%
        {braun2019fake}
\bibfield{author}{\bibinfo{person}{Joshua~A Braun} {and}
  \bibinfo{person}{Jessica~L Eklund}.} \bibinfo{year}{2019}\natexlab{}.
\newblock \showarticletitle{Fake News, Real Money: Ad Tech Platforms,
  Profit-Driven Hoaxes, and the Business of Journalism}.
\newblock \bibinfo{journal}{\emph{Digital Journalism}} \bibinfo{volume}{7},
  \bibinfo{number}{1} (\bibinfo{year}{2019}), \bibinfo{pages}{1--21}.
\newblock


\bibitem[\protect\citeauthoryear{Castelo, Almeida, Elghafari, Santos, Pham,
  Nakamura, and Freire}{Castelo et~al\mbox{.}}{2019}]%
        {castelo2019topic}
\bibfield{author}{\bibinfo{person}{Sonia Castelo}, \bibinfo{person}{Thais
  Almeida}, \bibinfo{person}{Anas Elghafari}, \bibinfo{person}{A{\'e}cio
  Santos}, \bibinfo{person}{Kien Pham}, \bibinfo{person}{Eduardo Nakamura},
  {and} \bibinfo{person}{Juliana Freire}.} \bibinfo{year}{2019}\natexlab{}.
\newblock \showarticletitle{A Topic-Agnostic Approach for Identifying Fake News
  Pages}. In \bibinfo{booktitle}{\emph{WWW Companion}}. ACM,
  \bibinfo{pages}{975--980}.
\newblock


\bibitem[\protect\citeauthoryear{Castillo, Mendoza, and Poblete}{Castillo
  et~al\mbox{.}}{2011}]%
        {castillo2011information}
\bibfield{author}{\bibinfo{person}{Carlos Castillo}, \bibinfo{person}{Marcelo
  Mendoza}, {and} \bibinfo{person}{Barbara Poblete}.}
  \bibinfo{year}{2011}\natexlab{}.
\newblock \showarticletitle{Information credibility on twitter}. In
  \bibinfo{booktitle}{\emph{Proceedings of the 20th WWW}}. ACM,
  \bibinfo{pages}{675--684}.
\newblock


\bibitem[\protect\citeauthoryear{Cruz, Rocha, Sousa-Silva, and Cardoso}{Cruz
  et~al\mbox{.}}{2019}]%
        {cruz2019team}
\bibfield{author}{\bibinfo{person}{Andr{\'e} Cruz}, \bibinfo{person}{Gil
  Rocha}, \bibinfo{person}{Rui Sousa-Silva}, {and}
  \bibinfo{person}{Henrique~Lopes Cardoso}.} \bibinfo{year}{2019}\natexlab{}.
\newblock \showarticletitle{Team Fernando-Pessa at SemEval-2019 Task 4: Back to
  Basics in Hyperpartisan News Detection}. In
  \bibinfo{booktitle}{\emph{Proceedings of the 13th SemEval}}.
  \bibinfo{pages}{999--1003}.
\newblock


\bibitem[\protect\citeauthoryear{F{\"a}rber, Qurdina, and Ahmedi}{F{\"a}rber
  et~al\mbox{.}}{2019}]%
        {farber2019team}
\bibfield{author}{\bibinfo{person}{Michael F{\"a}rber}, \bibinfo{person}{Agon
  Qurdina}, {and} \bibinfo{person}{Lule Ahmedi}.}
  \bibinfo{year}{2019}\natexlab{}.
\newblock \showarticletitle{Team Peter Brinkmann at SemEval-2019 Task 4:
  Detecting Biased News Articles Using Convolutional Neural Networks}. In
  \bibinfo{booktitle}{\emph{Proceedings of the 13th SemEval}}.
  \bibinfo{pages}{1032--1036}.
\newblock


\bibitem[\protect\citeauthoryear{Graham, Haidt, Koleva, Motyl, Iyer, Wojcik,
  and Ditto}{Graham et~al\mbox{.}}{2013}]%
        {graham2013moral}
\bibfield{author}{\bibinfo{person}{Jesse Graham}, \bibinfo{person}{Jonathan
  Haidt}, \bibinfo{person}{Sena Koleva}, \bibinfo{person}{Matt Motyl},
  \bibinfo{person}{Ravi Iyer}, \bibinfo{person}{Sean~P Wojcik}, {and}
  \bibinfo{person}{Peter~H Ditto}.} \bibinfo{year}{2013}\natexlab{}.
\newblock \showarticletitle{Moral foundations theory: The pragmatic validity of
  moral pluralism}.
\newblock In \bibinfo{booktitle}{\emph{Advances in experimental social
  psychology}}. Vol.~\bibinfo{volume}{47}. \bibinfo{publisher}{Elsevier},
  \bibinfo{pages}{55--130}.
\newblock


\bibitem[\protect\citeauthoryear{Grover and Leskovec}{Grover and
  Leskovec}{2016}]%
        {grover2016node2vec}
\bibfield{author}{\bibinfo{person}{Aditya Grover} {and} \bibinfo{person}{Jure
  Leskovec}.} \bibinfo{year}{2016}\natexlab{}.
\newblock \showarticletitle{node2vec: Scalable feature learning for networks}.
  In \bibinfo{booktitle}{\emph{Proceedings of the 22nd ACM SIGKDD international
  conference on Knowledge discovery and data mining}}. ACM,
  \bibinfo{pages}{855--864}.
\newblock


\bibitem[\protect\citeauthoryear{Guacho, Abdali, Shah, and Papalexakis}{Guacho
  et~al\mbox{.}}{2018}]%
        {guacho2018semi}
\bibfield{author}{\bibinfo{person}{Gisel~Bastidas Guacho},
  \bibinfo{person}{Sara Abdali}, \bibinfo{person}{Neil Shah}, {and}
  \bibinfo{person}{Evangelos~E Papalexakis}.} \bibinfo{year}{2018}\natexlab{}.
\newblock \showarticletitle{Semi-supervised Content-based Detection of
  Misinformation via Tensor Embeddings}. In \bibinfo{booktitle}{\emph{2018
  IEEE/ACM International Conference on Advances in Social Networks Analysis and
  Mining (ASONAM)}}. IEEE, \bibinfo{pages}{322--325}.
\newblock


\bibitem[\protect\citeauthoryear{Horne, Dron, Khedr, and Adal{\i}}{Horne
  et~al\mbox{.}}{2018}]%
        {horne2018accessing}
\bibfield{author}{\bibinfo{person}{Benjamin~D Horne}, \bibinfo{person}{William
  Dron}, \bibinfo{person}{Sara Khedr}, {and} \bibinfo{person}{Sibel Adal{\i}}.}
  \bibinfo{year}{2018}\natexlab{}.
\newblock \showarticletitle{Assessing the News Landscape: A Multi-Module
  Toolkit for Evaluating the Credibility of News}. In
  \bibinfo{booktitle}{\emph{WWW Companion}}.
\newblock


\bibitem[\protect\citeauthoryear{Horne, Gruppi, and Adal{\i}}{Horne
  et~al\mbox{.}}{2020}]%
        {horne2020all}
\bibfield{author}{\bibinfo{person}{Benjamin~D Horne},
  \bibinfo{person}{Maur{\'\i}cio Gruppi}, {and} \bibinfo{person}{Sibel
  Adal{\i}}.} \bibinfo{year}{2020}\natexlab{}.
\newblock \showarticletitle{Do All Good Actors Look The Same? Exploring News
  Veracity Detection Across The US and The UK}.
\newblock \bibinfo{journal}{\emph{ICWSM ICWSM Data Challenge on Safety}}
  (\bibinfo{year}{2020}).
\newblock


\bibitem[\protect\citeauthoryear{Horne, N{\o}rregaard, and Adal{\i}}{Horne
  et~al\mbox{.}}{2019a}]%
        {horne2019different}
\bibfield{author}{\bibinfo{person}{Benjamin~D Horne}, \bibinfo{person}{Jeppe
  N{\o}rregaard}, {and} \bibinfo{person}{Sibel Adal{\i}}.}
  \bibinfo{year}{2019}\natexlab{a}.
\newblock \showarticletitle{Different Spirals of Sameness: A Study of Content
  Sharing in Mainstream and Alternative Media}. In
  \bibinfo{booktitle}{\emph{Proceedings of ICWSM}}, Vol.~\bibinfo{volume}{13}.
  \bibinfo{pages}{257--266}.
\newblock


\bibitem[\protect\citeauthoryear{Horne, N{\o}rregaard, and Adali}{Horne
  et~al\mbox{.}}{2019b}]%
        {horne2019robust}
\bibfield{author}{\bibinfo{person}{Benjamin~D Horne}, \bibinfo{person}{Jeppe
  N{\o}rregaard}, {and} \bibinfo{person}{Sibel Adali}.}
  \bibinfo{year}{2019}\natexlab{b}.
\newblock \showarticletitle{Robust Fake News Detection Over Time and Attack}.
\newblock \bibinfo{journal}{\emph{ACM Transactions on Intelligent Systems and
  Technology (TIST)}} \bibinfo{volume}{11}, \bibinfo{number}{1}
  (\bibinfo{year}{2019}), \bibinfo{pages}{1--23}.
\newblock


\bibitem[\protect\citeauthoryear{Hosseinimotlagh and
  Papalexakis}{Hosseinimotlagh and Papalexakis}{2018}]%
        {hosseinimotlagh2018unsupervised}
\bibfield{author}{\bibinfo{person}{Seyedmehdi Hosseinimotlagh} {and}
  \bibinfo{person}{Evangelos~E Papalexakis}.} \bibinfo{year}{2018}\natexlab{}.
\newblock \showarticletitle{Unsupervised content-based identification of fake
  news articles with tensor decomposition ensembles}.
\newblock \bibinfo{journal}{\emph{MIS2}} (\bibinfo{year}{2018}).
\newblock


\bibitem[\protect\citeauthoryear{Hutto and Gilbert}{Hutto and Gilbert}{2014}]%
        {hutto2014vader}
\bibfield{author}{\bibinfo{person}{Clayton~J Hutto} {and} \bibinfo{person}{Eric
  Gilbert}.} \bibinfo{year}{2014}\natexlab{}.
\newblock \showarticletitle{Vader: A parsimonious rule-based model for
  sentiment analysis of social media text}. In \bibinfo{booktitle}{\emph{Eighth
  ICWSM}}.
\newblock


\bibitem[\protect\citeauthoryear{Kumar and Shah}{Kumar and Shah}{2018}]%
        {kumar2018false}
\bibfield{author}{\bibinfo{person}{Srijan Kumar} {and} \bibinfo{person}{Neil
  Shah}.} \bibinfo{year}{2018}\natexlab{}.
\newblock \showarticletitle{False information on web and social media: A
  survey}.
\newblock \bibinfo{journal}{\emph{arXiv preprint arXiv:1804.08559}}
  (\bibinfo{year}{2018}).
\newblock


\bibitem[\protect\citeauthoryear{Lazer, Baum, Benkler, Berinsky, Greenhill,
  Menczer, Metzger, Nyhan, Pennycook, Rothschild, et~al\mbox{.}}{Lazer
  et~al\mbox{.}}{2018}]%
        {lazer2018science}
\bibfield{author}{\bibinfo{person}{David~MJ Lazer}, \bibinfo{person}{Matthew~A
  Baum}, \bibinfo{person}{Yochai Benkler}, \bibinfo{person}{Adam~J Berinsky},
  \bibinfo{person}{Kelly~M Greenhill}, \bibinfo{person}{Filippo Menczer},
  \bibinfo{person}{Miriam~J Metzger}, \bibinfo{person}{Brendan Nyhan},
  \bibinfo{person}{Gordon Pennycook}, \bibinfo{person}{David Rothschild},
  {et~al\mbox{.}}} \bibinfo{year}{2018}\natexlab{}.
\newblock \showarticletitle{The science of fake news}.
\newblock \bibinfo{journal}{\emph{Science}} \bibinfo{volume}{359},
  \bibinfo{number}{6380} (\bibinfo{year}{2018}), \bibinfo{pages}{1094--1096}.
\newblock


\bibitem[\protect\citeauthoryear{Lewandowsky, Ecker, Seifert, Schwarz, and
  Cook}{Lewandowsky et~al\mbox{.}}{2012}]%
        {lewandowsky2012misinformation}
\bibfield{author}{\bibinfo{person}{Stephan Lewandowsky},
  \bibinfo{person}{Ullrich~KH Ecker}, \bibinfo{person}{Colleen~M Seifert},
  \bibinfo{person}{Norbert Schwarz}, {and} \bibinfo{person}{John Cook}.}
  \bibinfo{year}{2012}\natexlab{}.
\newblock \showarticletitle{Misinformation and its correction: Continued
  influence and successful debiasing}.
\newblock \bibinfo{journal}{\emph{Psychological Science in the Public
  Interest}} \bibinfo{volume}{13}, \bibinfo{number}{3} (\bibinfo{year}{2012}).
\newblock


\bibitem[\protect\citeauthoryear{Li and Goldwasser}{Li and Goldwasser}{2019}]%
        {li2019encoding}
\bibfield{author}{\bibinfo{person}{Chang Li} {and} \bibinfo{person}{Dan
  Goldwasser}.} \bibinfo{year}{2019}\natexlab{}.
\newblock \showarticletitle{Encoding Social Information with Graph
  Convolutional Networks forPolitical Perspective Detection in News Media}. In
  \bibinfo{booktitle}{\emph{Proceedings of ACL}}. \bibinfo{pages}{2594--2604}.
\newblock


\bibitem[\protect\citeauthoryear{Moreno, Pitarch, Pinel-Sauvagnat, and
  Hubert}{Moreno et~al\mbox{.}}{2019}]%
        {moreno2019rouletabille}
\bibfield{author}{\bibinfo{person}{Jose~G Moreno}, \bibinfo{person}{Yoann
  Pitarch}, \bibinfo{person}{Karen Pinel-Sauvagnat}, {and}
  \bibinfo{person}{Gilles Hubert}.} \bibinfo{year}{2019}\natexlab{}.
\newblock \showarticletitle{Rouletabille at SemEval-2019 Task 4: Neural Network
  Baseline for Identification of Hyperpartisan Publishers}. In
  \bibinfo{booktitle}{\emph{Proceedings of the 13th SemEval}}.
  \bibinfo{pages}{981--984}.
\newblock


\bibitem[\protect\citeauthoryear{Newman and Girvan}{Newman and Girvan}{2004}]%
        {newman2004finding}
\bibfield{author}{\bibinfo{person}{Mark~EJ Newman} {and}
  \bibinfo{person}{Michelle Girvan}.} \bibinfo{year}{2004}\natexlab{}.
\newblock \showarticletitle{Finding and evaluating community structure in
  networks}.
\newblock \bibinfo{journal}{\emph{Physical review E}} \bibinfo{volume}{69},
  \bibinfo{number}{2} (\bibinfo{year}{2004}), \bibinfo{pages}{026113}.
\newblock


\bibitem[\protect\citeauthoryear{N{\o}rregaard, Horne, and
  Adal{\i}}{N{\o}rregaard et~al\mbox{.}}{2019}]%
        {norregaard2019nela}
\bibfield{author}{\bibinfo{person}{Jeppe N{\o}rregaard},
  \bibinfo{person}{Benjamin~D Horne}, {and} \bibinfo{person}{Sibel Adal{\i}}.}
  \bibinfo{year}{2019}\natexlab{}.
\newblock \showarticletitle{NELA-GT-2018: A Large Multi-Labelled News Dataset
  for the Study of Misinformation in News Articles}. In
  \bibinfo{booktitle}{\emph{Proceedings ofICWSM}}, Vol.~\bibinfo{volume}{13}.
  \bibinfo{pages}{630--638}.
\newblock


\bibitem[\protect\citeauthoryear{Pennycook, McPhetres, Zhang, Lu, and
  Rand}{Pennycook et~al\mbox{.}}{2020}]%
        {pennycook2020fighting}
\bibfield{author}{\bibinfo{person}{Gordon Pennycook}, \bibinfo{person}{Jonathon
  McPhetres}, \bibinfo{person}{Yunhao Zhang}, \bibinfo{person}{Jackson~G Lu},
  {and} \bibinfo{person}{David~G Rand}.} \bibinfo{year}{2020}\natexlab{}.
\newblock \showarticletitle{Fighting COVID-19 misinformation on social media:
  Experimental evidence for a scalable accuracy-nudge intervention}.
\newblock \bibinfo{journal}{\emph{Psychological science}} \bibinfo{volume}{31},
  \bibinfo{number}{7} (\bibinfo{year}{2020}), \bibinfo{pages}{770--780}.
\newblock


\bibitem[\protect\citeauthoryear{Popat, Mukherjee, Str{\"o}tgen, and
  Weikum}{Popat et~al\mbox{.}}{2016}]%
        {popat2016credibility}
\bibfield{author}{\bibinfo{person}{Kashyap Popat}, \bibinfo{person}{Subhabrata
  Mukherjee}, \bibinfo{person}{Jannik Str{\"o}tgen}, {and}
  \bibinfo{person}{Gerhard Weikum}.} \bibinfo{year}{2016}\natexlab{}.
\newblock \showarticletitle{Credibility assessment of textual claims on the
  web}. In \bibinfo{booktitle}{\emph{Proceedings of ACM CIKM}}. ACM,
  \bibinfo{pages}{2173--2178}.
\newblock


\bibitem[\protect\citeauthoryear{Potthast, Kiesel, Reinartz, Bevendorff, and
  Stein}{Potthast et~al\mbox{.}}{2017}]%
        {potthast2017stylometric}
\bibfield{author}{\bibinfo{person}{Martin Potthast}, \bibinfo{person}{Johannes
  Kiesel}, \bibinfo{person}{Kevin Reinartz}, \bibinfo{person}{Janek
  Bevendorff}, {and} \bibinfo{person}{Benno Stein}.}
  \bibinfo{year}{2017}\natexlab{}.
\newblock \showarticletitle{A Stylometric Inquiry into Hyperpartisan and Fake
  News}.
\newblock \bibinfo{journal}{\emph{arXiv preprint arXiv:1702.05638}}
  (\bibinfo{year}{2017}).
\newblock


\bibitem[\protect\citeauthoryear{Recasens, Danescu-Niculescu-Mizil, and
  Jurafsky}{Recasens et~al\mbox{.}}{2013}]%
        {recasens2013linguistic}
\bibfield{author}{\bibinfo{person}{Marta Recasens}, \bibinfo{person}{Cristian
  Danescu-Niculescu-Mizil}, {and} \bibinfo{person}{Dan Jurafsky}.}
  \bibinfo{year}{2013}\natexlab{}.
\newblock \showarticletitle{Linguistic models for analyzing and detecting
  biased language}. In \bibinfo{booktitle}{\emph{Proceedings of the 51st Annual
  Meeting of the ACL (Volume 1: Long Papers)}}, Vol.~\bibinfo{volume}{1}.
  \bibinfo{pages}{1650--1659}.
\newblock


\bibitem[\protect\citeauthoryear{Rozemberczki, Allen, and Sarkar}{Rozemberczki
  et~al\mbox{.}}{2019}]%
        {rozemberczki2019multi}
\bibfield{author}{\bibinfo{person}{Benedek Rozemberczki}, \bibinfo{person}{Carl
  Allen}, {and} \bibinfo{person}{Rik Sarkar}.} \bibinfo{year}{2019}\natexlab{}.
\newblock \showarticletitle{Multi-scale Attributed Node Embedding}.
\newblock \bibinfo{journal}{\emph{arXiv preprint arXiv:1909.13021}}
  (\bibinfo{year}{2019}).
\newblock


\bibitem[\protect\citeauthoryear{Ruchansky, Seo, and Liu}{Ruchansky
  et~al\mbox{.}}{2017}]%
        {ruchansky2017csi}
\bibfield{author}{\bibinfo{person}{Natali Ruchansky}, \bibinfo{person}{Sungyong
  Seo}, {and} \bibinfo{person}{Yan Liu}.} \bibinfo{year}{2017}\natexlab{}.
\newblock \showarticletitle{Csi: A hybrid deep model for fake news detection}.
  In \bibinfo{booktitle}{\emph{Proceedings of the 2017 ACM on Conference on
  Information and Knowledge Management}}. ACM, \bibinfo{pages}{797--806}.
\newblock


\bibitem[\protect\citeauthoryear{Singh, Bansal, Bode, Budak, Chi, Kawintiranon,
  Padden, Vanarsdall, Vraga, and Wang}{Singh et~al\mbox{.}}{2020}]%
        {singh2020first}
\bibfield{author}{\bibinfo{person}{Lisa Singh}, \bibinfo{person}{Shweta
  Bansal}, \bibinfo{person}{Leticia Bode}, \bibinfo{person}{Ceren Budak},
  \bibinfo{person}{Guangqing Chi}, \bibinfo{person}{Kornraphop Kawintiranon},
  \bibinfo{person}{Colton Padden}, \bibinfo{person}{Rebecca Vanarsdall},
  \bibinfo{person}{Emily Vraga}, {and} \bibinfo{person}{Yanchen Wang}.}
  \bibinfo{year}{2020}\natexlab{}.
\newblock \showarticletitle{A first look at COVID-19 information and
  misinformation sharing on Twitter}.
\newblock \bibinfo{journal}{\emph{arXiv preprint arXiv:2003.13907}}
  (\bibinfo{year}{2020}).
\newblock


\bibitem[\protect\citeauthoryear{Singhania, Fernandez, and Rao}{Singhania
  et~al\mbox{.}}{2017}]%
        {singhania20173han}
\bibfield{author}{\bibinfo{person}{Sneha Singhania}, \bibinfo{person}{Nigel
  Fernandez}, {and} \bibinfo{person}{Shrisha Rao}.}
  \bibinfo{year}{2017}\natexlab{}.
\newblock \showarticletitle{3HAN: A Deep Neural Network for Fake News
  Detection}. In \bibinfo{booktitle}{\emph{NeuIPS}}. Springer,
  \bibinfo{pages}{572--581}.
\newblock


\bibitem[\protect\citeauthoryear{Speed and Mannion}{Speed and Mannion}{2017}]%
        {speed2017rise}
\bibfield{author}{\bibinfo{person}{Ewen Speed} {and} \bibinfo{person}{Russell
  Mannion}.} \bibinfo{year}{2017}\natexlab{}.
\newblock \showarticletitle{The rise of post-truth populism in pluralist
  liberal democracies: challenges for health policy}.
\newblock \bibinfo{journal}{\emph{International journal of health policy and
  management}} \bibinfo{volume}{6}, \bibinfo{number}{5} (\bibinfo{year}{2017}),
  \bibinfo{pages}{249}.
\newblock


\bibitem[\protect\citeauthoryear{Starbird}{Starbird}{2017}]%
        {starbird2017examining}
\bibfield{author}{\bibinfo{person}{Kate Starbird}.}
  \bibinfo{year}{2017}\natexlab{}.
\newblock \showarticletitle{Examining the Alternative Media Ecosystem Through
  the Production of Alternative Narratives of Mass Shooting Events on
  Twitter.}. In \bibinfo{booktitle}{\emph{ICWSM}}. \bibinfo{pages}{230--239}.
\newblock


\bibitem[\protect\citeauthoryear{Starbird, Arif, Wilson, Van~Koevering,
  Yefimova, and Scarnecchia}{Starbird et~al\mbox{.}}{2018}]%
        {starbird2018ecosystem}
\bibfield{author}{\bibinfo{person}{Kate Starbird}, \bibinfo{person}{Ahmer
  Arif}, \bibinfo{person}{Tom Wilson}, \bibinfo{person}{Katherine
  Van~Koevering}, \bibinfo{person}{Katya Yefimova}, {and}
  \bibinfo{person}{Daniel Scarnecchia}.} \bibinfo{year}{2018}\natexlab{}.
\newblock \showarticletitle{Ecosystem or Echo-System? Exploring Content Sharing
  across Alternative Media Domains}.
\newblock  (\bibinfo{year}{2018}).
\newblock


\bibitem[\protect\citeauthoryear{Tausczik and Pennebaker}{Tausczik and
  Pennebaker}{2010}]%
        {tausczik2010psychological}
\bibfield{author}{\bibinfo{person}{Yla~R Tausczik} {and}
  \bibinfo{person}{James~W Pennebaker}.} \bibinfo{year}{2010}\natexlab{}.
\newblock \showarticletitle{The psychological meaning of words: LIWC and
  computerized text analysis methods}.
\newblock \bibinfo{journal}{\emph{Journal of language and social psychology}}
  \bibinfo{volume}{29}, \bibinfo{number}{1} (\bibinfo{year}{2010}),
  \bibinfo{pages}{24--54}.
\newblock


\bibitem[\protect\citeauthoryear{Yang, Liu, Zhao, Sun, and Chang}{Yang
  et~al\mbox{.}}{2015}]%
        {yang2015network}
\bibfield{author}{\bibinfo{person}{Cheng Yang}, \bibinfo{person}{Zhiyuan Liu},
  \bibinfo{person}{Deli Zhao}, \bibinfo{person}{Maosong Sun}, {and}
  \bibinfo{person}{Edward Chang}.} \bibinfo{year}{2015}\natexlab{}.
\newblock \showarticletitle{Network representation learning with rich text
  information}. In \bibinfo{booktitle}{\emph{Twenty-Fourth IJCAI}}.
\newblock


\bibitem[\protect\citeauthoryear{Yang, Liu, Yu, and Yang}{Yang
  et~al\mbox{.}}{2012}]%
        {yang2012automatic}
\bibfield{author}{\bibinfo{person}{Fan Yang}, \bibinfo{person}{Yang Liu},
  \bibinfo{person}{Xiaohui Yu}, {and} \bibinfo{person}{Min Yang}.}
  \bibinfo{year}{2012}\natexlab{}.
\newblock \showarticletitle{Automatic detection of rumor on Sina Weibo}. In
  \bibinfo{booktitle}{\emph{Proceedings of the ACM SIGKDD Workshop on Mining
  Data Semantics}}. \bibinfo{pages}{13}.
\newblock


\bibitem[\protect\citeauthoryear{Ye and Skiena}{Ye and Skiena}{2019}]%
        {ye2019mediarank}
\bibfield{author}{\bibinfo{person}{Junting Ye} {and} \bibinfo{person}{Steven
  Skiena}.} \bibinfo{year}{2019}\natexlab{}.
\newblock \showarticletitle{MediaRank: Computational Ranking of Online News
  Sources}.
\newblock \bibinfo{journal}{\emph{arXiv preprint arXiv:1903.07581}}
  (\bibinfo{year}{2019}).
\newblock


\end{thebibliography}
\end{document}